\documentclass{article}
\usepackage{cite}
\usepackage{graphicx}

\begin{document} 

\newcommand{\bear}{\begin{eqnarray}}
\newcommand{\eear}{\end{eqnarray}}
\newcommand{\be}{\begin{equation}}
\newcommand{\ee}{\end{equation}}
\newcommand{\beqn}{\begin{eqnarray}}
\newcommand{\eeqn}{\end{eqnarray}}
\newcommand{\beqnn}{\begin{eqnarray*}}
\newcommand{\eeqnn}{\end{eqnarray*}}

\def\vep{\varepsilon}
\def\vf{\varphi}

\begin{center} {\Large \bf
Adiabatic versus instantaneous transitions from a harmonic oscillator to an inverted oscillator}

\end{center}

\begin{center} {\bf
Viktor V. Dodonov
and Alexandre V. Dodonov }
\end{center}
 
\begin{center}

{\it 
Institute of Physics and International Center of Physics, University of Brasilia,
Brasilia, DF, Brazil }

\end{center}

{Correspondence: vicdod@gmail.com, adodonov@unb.br}

\section{Introduction}

We consider a quantum system described by means of one of the simplest examples of Hamiltonians
with time-dependent parameters:
\be
H = p^2/2 + \gamma(t) x^2/2.
\label{Ham}
\ee
For the sake of simplicity, we assume the unit value of particle's mass.
One of many brilliant results of classical and quantum mechanics is the existence of 
{\em adiabatic invariants\/} in the case when parameters of a system vary slowly with time.
If $\gamma(t) = \omega^2(t)$ with a {\em real\/}  frequency $\omega(t)$,
the well known adiabatic invariant is the ratio of the energy of a harmonic oscillator ${\cal E}(t)$ to its 
time-dependent frequency $\omega(t)$ \cite{Land-mech}:
\be 
{\cal E}(t)/ \omega(t) = const .
\label{inv}
\ee
The condition of validity of Eq. (\ref{inv}) can be written as 
\be
 \dot\gamma ^2/|\gamma(t)|^3 \ll 1.
\label{condval}
\ee
This condition is obviously broken, if function $\gamma(t)$ passes through zero value. 
Recently \cite{aditrip}, we studied the case when this function always remains non-negative, even 
after attaining zero values.
In particular, it was shown that formula (\ref{inv}) remains valid after the frequency crosses zero
value, but the value of the constant in the right-hand side is different from the value before the
crossing. If function $\gamma(t)$ behaves as some power function ($\sim t^n$) in the vicinity of time
instant when $\gamma(t)=0$, this new value depends on the power $n$ in the law of crossing zero.

Now, our main goal was to study, what can happen 
if function $\gamma(t)$ slowly passes through zero value and becomes {\em negative}. 
This means that the initial {\em harmonic oscillator\/} becomes an {\em inverted oscillator}.
Note that Schwabl and Thirring \cite{Schwabl64} and especially Glauber \cite{Glaub-inv,Glaub86} 
used the Hamiltonian 
$H = -p^2/2 - \omega^2 x^2/2$ 
(or $H = -\hbar\omega a^{\dagger}a$) as a simple
model of a quantum amplifier. In order to avoid a confusion, we call this model as a 
``totally inverted oscillator''. We consider here the ``partially inverted oscillator'', with the
positive term $p^2/2$ in the Hamiltonian, using a short name ``inverted oscillator''.
 
The history of studies of the inverted oscillator in the quantum regime 
(called also sometimes as an ``upside-down oscillator'', or ``repulsive oscillator'', 
or ``reversed oscillator'') 
is surprisingly long
\cite{MoshSel78,DM79,Leach80,Guth85,Barton86,Scully86,Bal90,Papad90,DodNik91,Baskout94,Kleber94,Nogami94,%
Zurek94,Alb94,Bhad97,Cast97,Nieto97,Papad97,Voros97,Adam98,Sarkar98,Isar00,Shim00,Mat00,Zur03,Choi04,Chrus04,%
Pedrosa04,Lombar05,Yuce06,Combesc07,Yeon08,ManMarStor08,Wolf10,KimKim10,Guo11a,Guo11,Lewis11,Klaud12,Rosu12,%
Schleich13,Bagrov13,Bermud13,Steu14,Gentil15,Pedrosa15,Schulze15,Maam16,Marc16,Rauh16,Yuce16,Finster17,%
MaamChoi17,Rajeev18,Chist19,Morita19,Mota19,Ali20,Aouda20,Golov20,Moya20,Nikitin20,Subram21,Bhat21,Gietka21,%
Hoj21,Yuce21,Bag22,Ullinger22,ManMan23,Maam23,Sacc23,Cher23} 
(the ``totally inverted oscillator'' was considered in \cite{Woger86,Tarzi88,Grimau20}).
However, it seems that the energy evolution in the adiabatic regime was not considered in all cited
papers. Another interesting (and exactly solvable) case corresponds to a sudden jump
of function $\gamma(t)$ at $t=0$ from the value $\omega_0^2$ to some constant negative value $-\kappa^2$.
An instantaneous frequency jump between two positive values 
of coefficient $\gamma$ was studied in numerous papers
\cite{Parker71,JanYu,Gra,Ma,Lo90,JanAd92,DKN93,Kiss94,Sas94,Mend00,Tiba21}.
However, the jump to the negative value of $\gamma$ was not considered before.
Therefore, we hope that the results of our study
can be interesting for many readers.


\section{Evolution of the mean oscillator energy }
\label{sec-quantgen}

We assume that function $\gamma(t)$ equals zero at $t=0$. Then, the initial conditions can be chosen
at some negative value $t=-\tau$ of the time variable. 
If the evolution is governed by Hamiltonian (\ref{Ham}),
 mean values of the coordinate and
momentum can be expressed in terms of two fundamental solutions of the classical equation
\be
\ddot\vep + \gamma(t)\vep = 0.
\label{eq}
\ee
We suppose that $\gamma(t) =\omega_0^2 = const$ for $t \le -\tau$, with $\omega_0 >0$.
If $\gamma(t)$ is a real function, it is convenient, following
the seminal papers \cite{Husimi,PP,LR,MMT70} (see also, e.g., a recent review \cite{Suslov10}), 
to choose the fundamental solutions 
in the form of mutually conjugated complex functions  $\vep(t)$ and $\vep^*(t)$, 
satisfying the initial conditions 
\be
\vep(-\tau)= \omega_{0}^{-1/2}, \quad \dot\vep(-\tau)= i\omega_{0}^{1/2},
\label{invep}
\ee 
so that the Wronskian identity for the solutions $\vep(t)$ and $\vep^*(t)$ has the form
\be
\dot\vep(t) \vep^*(t) - \dot\vep^*(t) \vep(t) \equiv 2i.
\label{Wr}
\ee
Then, we can write at $t\ge -\tau$ the following expressions for the first-order mean values
(which have the same form in the classical and quantum cases):
\be
x(t) = x_0 \sqrt{\omega_0}\, \mbox{Re}[\vep(t)] + \frac{p_0}{\sqrt{\omega_0}} \mbox{Im}[\vep(t)], \quad
p(t) = x_0 \sqrt{\omega_0}\,\mbox{Re}[\dot\vep(t)] + \frac{p_0}{\sqrt{\omega_0}} \mbox{Im}[\dot\vep(t)],
\label{solxp}
\ee
where $x_0=x(-\tau)$ and $p_0=p(-\tau)$.

The second-order moments of the canonical operators evolve at $t \ge -\tau$ as follows:
\be
\langle x^2\rangle_t =  \langle x^2\rangle_{0}\, \omega_0 \left(\mbox{Re}[\vep_t]\right)^2 + 
\frac{\langle p^2\rangle_{0}}{\omega_0} \left(\mbox{Im}[\vep_t]\right)^2
+ \langle xp + px \rangle_{0} \left(\mbox{Re}[\vep_t]\right) \left(\mbox{Im}[\vep_t]\right),
\label{x2t}
\ee
\be
\langle p^2\rangle_t =  \langle x^2\rangle_{0}\, \omega_0 \left(\mbox{Re}[\dot\vep_t]\right)^2 + 
\frac{\langle p^2\rangle_{0}}{\omega_0} \left(\mbox{Im}[\dot\vep_t]\right)^2
+ \langle xp + px \rangle_{0} \left(\mbox{Re}[\dot\vep_t]\right) \left(\mbox{Im}[\dot\vep_t]\right).
\label{p2t}
\ee
\be
\langle xp + px \rangle_t = 2 \langle x^2\rangle_{0}\, \omega_0 \mbox{Re}[\vep_t] \mbox{Re}[\dot\vep_t] + 
\frac{2\langle p^2\rangle_{0}}{\omega_0} \mbox{Im}[\vep_t] \mbox{Im}[\dot\vep_t]
+ \langle xp + px \rangle_{0} \mbox{Im}[\dot\vep_t\vep_t].
\label{pxt}
\ee
These expressions can be easily obtained, if one considers $x(t)$ and $p(t)$ as the Heisenberg operators
and calculates the average values of their powers and products over the initial state. Such an approach
works well for any quadratic Hamiltonian.

The time-dependent mean energy is given by the formula
\be
 {\cal E}(t) = \frac12\left[\langle p^2 \rangle_t 
+ \gamma(t) \langle x^2 \rangle_t \right].
\label{defE}
\ee
It is worth remembering that for systems with {\em quadratic\/} Hamiltonians with respect to $x$
and $p$, the dynamics of the first-order mean values $\langle x\rangle$ and $\langle p\rangle$
are {\em totally independent\/} from the dynamics of the variances 
$\sigma_x = \langle x^2\rangle - \langle x\rangle^2$, 
$\sigma_p = \langle p^2\rangle - \langle p\rangle^2$ and
$\sigma_{xp} = \langle xp +px\rangle/2 - \langle x\rangle \langle p\rangle$.
This means that the equations of the same form as (\ref{x2t}) and (\ref{p2t}) exist for the
sets $(\langle x\rangle^2, \langle p\rangle^2, \langle x\rangle \langle p\rangle)$ and
$(\sigma_x, \sigma_p, \sigma_{xp})$.

\subsection{Adiabatic evolution if the oscillator remains harmonic}

The adiabatic (quasiclassical) approximate complex solution to Eq. (\ref{eq}),
satisfying the initial conditions (\ref{invep}),
has the following form at $t <0$ [when $\gamma(t) >0$ and $\omega(t) >0$]:
\be
\vep(t) \approx [\omega(t)]^{-1/2} e^{i\phi_{\tau}(t)}, 
\qquad \dot\vep(t) \approx i[\omega(t)]^{1/2} e^{i\phi_{\tau}(t)}, 
\qquad \phi_{\tau}(t) = \int_{-\tau}^t\omega(z)dz.
\label{adsol}
\ee
Putting the solution (\ref{adsol}) in the equations (\ref{x2t})-(\ref{defE}).
 we arrive immediately at the adiabatic invariant 
\be
{\cal E}(t)/\omega(t) = {\cal E}(-\tau)/\omega_0 , 
\label{adiinv}
\ee
{\em for arbitrary initial values\/} at $t=-\tau$.

The solution (\ref{adsol}) obviously looses its sense if $\omega(t) =0$ 
at some time instant $t_*$ (taken as $t=0$ in our paper).
However, if the function $\gamma(t)$ slowly passes through zero value and becomes positive 
(and not too small) again, 
the condition of validity of the quasiclassical approximation (\ref{condval}) is reestablished again. 
If $\gamma(t) = \omega^2(t) >0$ for $t>0$, 
the solution for $t>0$ (outside some interval near $t=0$) can be written as \cite{aditrip}
 \be
\vep(t) \approx [\omega(t)]^{-1/2} \left[ u_{+} e^{i{\phi}(t)} + 
 u_{-} e^{-i{\phi}(t)} \right], 
\qquad
\dot\vep(t) \approx i[\omega(t)]^{1/2} 
\left[ u_{+} e^{i{\phi}(t)} -  u_{-} e^{-i{\phi}(t)} \right],
\label{adsol+}
\ee
where
\be
\phi(t) = \int_{0}^t\omega(z)dz, \qquad d\phi(t)/dt =\omega(t).
\label{t*}
\ee
Constant complex coefficients $u_{\pm}$ must obey the condition  
\be
|u_{+}|^2 - |u_{-}|^2 =1,
\label{uvcond}
\ee
which is the consequence of Eq. (\ref{Wr}). 
Then, Eq. (\ref{defE}) assumes the form \cite{aditrip}
\be
\frac{\langle {\cal E}\rangle_t }{\langle {\cal E}\rangle_{-\tau}} = 
 \frac{\omega(t)}{\omega_0} (\beta + \Delta\beta), 
\label{E-u-}
\ee
where
\be
 \beta = |u_{+}|^2 + |u_{-}|^2 = 1 + 2|u_{-}|^2,
\label{defbeta}
\ee
\be
\Delta\beta = \left\{ \left[\omega_0^2 \langle x^2 \rangle_{-\tau} - \langle p^2 \rangle_{-\tau}
\right]  \mbox{Re}\left(u_{+} u_{-}\right) +
\omega_0 \langle xp + px \rangle_{-\tau} \mbox{Im}\left(u_{+} u_{-}\right) \right\}
/{\langle {\cal E}\rangle_{-\tau}}.
\label{deltabeta}
\ee

Eq. (\ref{E-u-}) can be interpreted as a generalized adiabatic formula for the energy 
after the frequency passes {\em slowly\/} through zero value and remains real. 
It shows that the quantum mechanical mean energy
is proportional to the instant frequency $\omega(t)$ in the adiabatic regime. 
However, the proportionality coefficient
strongly depends on the initial conditions in the most general case. 
For this reason, we pay special attention to the case when 
\be
\langle p^2\rangle_{-\tau} = \omega_0^2 \langle x^2\rangle_{-\tau}, \qquad 
\langle xp + px \rangle_{-\tau}=0.
\label{special}
\ee
It includes the vacuum, thermal and Fock initial quantum states.
Then, $\Delta\beta =0$. In addition, many formulas can be simplified:
\[
\langle x^2\rangle_t = \omega_0 \langle x^2\rangle_{-\tau} |\vep(t)|^2, \qquad
\langle p^2 \rangle_t  = \langle p^2\rangle_{-\tau} |\dot\vep(t)|^2/\omega_0, 
\qquad
\langle xp + px \rangle_t = 2 \langle x^2\rangle_{-\tau}\, \omega_0 \mbox{Re}[\dot\vep(t)\vep^*(t)] ,
\]
\be
\langle {\cal E}\rangle_t = \frac{\langle {\cal E}\rangle_{-\tau}}{2\omega_0} 
\left[ \gamma(t) |\vep(t)|^2 + |\dot\vep(t)|^2\right].
\label{Etvac}
\ee
Note that the combination 
\be
D \equiv \langle x^2\rangle \langle p^2 \rangle - \langle xp + px \rangle^2/4
\ee
does not depend on time due to identity (\ref{Wr}). This is the simplest example of
{\em universal quantum invariants}, which hold for all systems with quadratic Hamiltonians \cite{Doduniv}.

In the adiabatic regime, the second-order moments (\ref{x2t})-(\ref{pxt}) can be written in terms of 
coefficients $u_{\pm}$ as follows [in the case of the special initial conditions (\ref{special})],
\[
\langle x^2\rangle_t =  \langle x^2\rangle_{-\tau} \frac{\omega_0}{\omega(t)} 
\left[ |u_{+}|^2 + |u_{-}|^2 + 2\mbox{Re} \left(u_{+}u_{-}^* e^{2i\phi(t)} \right)\right],
 \]
\[
\langle p^2 \rangle_t  = \langle p^2\rangle_{-\tau} \frac{\omega(t)}{\omega_0} 
\left[ |u_{+}|^2 + |u_{-}|^2 - 2\mbox{Re} \left(u_{+}u_{-}^* e^{2i\phi(t)} \right)\right],
 \]
\[
\langle xp + px \rangle_t = -4 \langle x^2\rangle_{-\tau}\, \omega_0 
\mbox{Im}\left(u_{+}u_{-}^* e^{2i\phi(t)} \right) .
\]
Defining the ``degree of squeezing'' $s$ as the ratio of the coordinate second-order moment to the 
variance $\hbar/(2\omega)$ in the ground state of the oscillator with the instant frequency $\omega$
(remember that we assume the unit mass),
we obtain the relations
\be
s_t/s_{-\tau} = |u_{+}|^2 + |u_{-}|^2 + 2\mbox{Re} \left(u_{+}u_{-}^* e^{2i\phi(t)} \right), 
\qquad
\left(|u_{+}| + |u_{-}|\right)^{-2} \le s_t/s_{-\tau} \le \left(|u_{+}| + |u_{-}|\right)^2.
\ee
Hence, the initial ground state ($s_{-\tau}=1$) can be squeezed when the frequency passes through zero
(remaining real).

\section{Exact solutions for the power profile of the frequency}
\label{sec-power}

The simplest function $\gamma(t)$ which changes its sign at the point $t=0$ is the linear one:
$\gamma(t) = - \omega_0^2 t/\tau$.
We consider a more general power dependence:
\be
\gamma(t) = \omega_0^2 \times \left\{
\begin{array}{cc}
[(-t)/\tau]^n, & t \le 0
\\
-[t/\tau]^n, & t \ge 0
\end{array}
\right. , \qquad n>0.
\label{gt}
\ee
For $t<0$, making the transformation
\be
\vep(t) = \sqrt{|t|}\,Z[y(t)], \qquad  y(t) = g \left|\frac{t}{\tau}\right|^{b}, 
\label{trans}
\ee
one can reduce Eq. (\ref{eq}) to the Bessel equation 
\be
\frac{d^2 Z}{dy^2} + \frac{1}{y}\frac{d Z}{dy} +\left(1- \frac{\nu^2}{y^2}\right) Z =0,
\label{eqBes}
\ee
with the following parameters:
\be
 \nu = \frac{1}{n+2} < \frac12,  \qquad
b = \frac{1}{2\nu}, \qquad g=2G\nu, \qquad G=\omega_0\tau.
\label{transZ}
\ee
For $t>0$, the same transformation (\ref{trans}) reduces Eq. (\ref{eq}) to the 
modified Bessel equation
\be
\frac{d^2 Z}{dy^2} + \frac{1}{y}\frac{d Z}{dy} -\left(1+ \frac{\nu^2}{y^2}\right) Z =0,
\label{eqmodBes}
\ee
with the same parameters (\ref{transZ}).

Hence, the function $\vep(t)$ can be written as a superposition of the usual Bessel functions
 $J_{\nu}(y)$ and $J_{-\nu}(y)$ for $t<0$, and a superposition of the modified Bessel functions
 $I_{\nu}(y)$ and $I_{-\nu}(y)$ for $t>0$:
\be
\vep(t) = \sqrt{|t|}\times \left\{
\begin{array}{ll}
\left\{A_{-} J_{\nu}[y(t)] + B_{-} J_{-\nu}[y(t)]\right\}, & t<0
\\
\left\{A_{+} I_{\nu}[y(t)] + B_{+} I_{-\nu}[y(t)]\right\}, & t>0
\end{array} \right. .
\label{vepBes}
\ee
Constant complex coefficients $A_{-}$ and $B_{-}$ can be found from the initial conditions (\ref{invep}).
Remembering that $d|t|/dt =-1$ for $t<0$, one obtains the following equations:
\[
A_{-} J_{\nu}\left(g \right) + B_{-} J_{-\nu}\left(g \right) = 1/\sqrt{G},
\qquad
A_{-} J^{\prime}_{\nu}\left(g \right) + B_{-} J^{\prime}_{-\nu}\left(g \right) 
=  -\,\frac{1}{\sqrt{G}} \left( i + \frac{1}{2G}\right),
\]
where $J^{\prime}_{\pm\nu}(z)$ means the derivative of the Bessel function $J_{\pm\nu}(z)$ with respect to its argument $z$.
Using the known Wronskian \cite{Grad,BE}
\[
J_{\nu}(z)J^{\prime}_{-\nu}(z) - J_{-\nu}(z)J^{\prime}_{\nu}(z) = -2\sin(\nu\pi)/(z\pi),
\]
one can obtain the following expressions:
\[
A_{-} = -\,\frac{\nu \pi \sqrt{G}}{\sin(\nu\pi)}\left[ J^{\prime}_{-\nu}\left(g \right)
+ \left( i + \frac{1}{2G}\right)J_{-\nu}\left(g \right) \right],
\qquad
B_{-} = \frac{\nu \pi \sqrt{G}}{\sin(\nu\pi)}\left[ 
 \left( i + \frac{1}{2G}\right)J_{\nu}\left(g \right)
 + J^{\prime}_{\nu}\left(g \right) \right].
\]
Using the known identities (see, e.g., formulas 7.2(54) and 7.2(55) in \cite{BE})
\be
J_{\nu}(z) \pm \frac{z}{\nu}J^{\prime}_{\nu}(z) = \frac{z}{\nu}J_{\nu \mp 1}(z),
\label{idenJnu}
\ee
one can simplify formulas for the coefficients $A_{-}$ and $B_{-}$:
\be
A_{-} = \frac{\nu \pi \sqrt{G}}{\sin(\nu\pi)}\left[ 
J_{1-\nu}\left(g \right) -i J_{-\nu}\left(g \right) 
\right],
\qquad
B_{-} = \frac{\nu \pi \sqrt{G}}{\sin(\nu\pi)}\left[ 
 i J_{\nu}\left(g \right)
 +J_{\nu -1}\left(g \right) \right].
 \label{A-B-}
 \ee

Coefficients $A_{+}$ and $B_{+}$ can be obtained from the conditions of the continuity of function
$\vep(t)$ and its time derivative at $t=0$.
The leading terms of the Bessel function $J_p(z)$ and the modified Bessel function $I_p(z)$ 
coincide at $z \to 0$:
\be
J_p(z) \approx I_p(z) \approx z^p/[2^p \Gamma(p+1)] , \qquad z \to 0.
\label{Jnuy0}
\ee
This means that   $\sqrt{|t|}J_{\nu}(y) \to 0$ and $\sqrt{|t|}I_{\nu}(y) \to 0$ when $t\to 0$.
On the other hand, the products $\sqrt{|t|}J_{-\nu}(y)$ and $\sqrt{|t|}I_{-\nu}(y)$ tend to 
identical finite values in this limit. 
Consequently, the continuity of function $\vep(t)$ 
at $t=0$ implies the condition $B_{+}=B_{-}$. 

The time derivative of function (\ref{vepBes}) at $t<0$ (when $d|t|/dt = -1$) can be written with the aid of identities  (\ref{idenJnu}) as follows:
\be
d\vep/dt = \frac{y}{2\nu\sqrt{|t|}}\left[  B_{-} J_{1-\nu}(y) - A_{-} J_{\nu-1}(y)\right], \qquad t\le 0.
\ee
On the other hand, using the special cases of Eqs. 7.11(19) and 7.11(20) from \cite{BE},
\be
I_{\nu}(z) \pm \frac{z}{\nu}I^{\prime}_{\nu}(z) = \pm\frac{z}{\nu}I_{\nu \mp 1}(z),
\label{idenInu}
\ee
we can write
\be
d\vep/dt = \frac{y}{2\nu\sqrt{t}}\left[  A_{+} I_{\nu-1}(y) +  B_{+} I_{1-\nu}(y) \right], \qquad t\ge 0.
\ee
Now, we notice that
$y J_{1-\nu}(y)/\sqrt{|t|} \to 0$ and $y I_{1-\nu}(y)/\sqrt{|t|} \to 0$ at $t\to 0$, while 
the products $yJ_{\nu-1}(y)/\sqrt{|t|}$ and $yI_{\nu-1}(y)/\sqrt{|t|}$ tend to identical 
finite values in this limit. 
Hence, the continuity of derivative $d\vep/dt$ at $t=0$ results in
 the condition $A_{+} = -A_{-}$.
Then, one can verify that the Wronskian identity (\ref{Wr}) 
is satisfied both for $t\le 0$ and $t\ge 0$, in view of the relations \cite{BE}
\be
J_{\nu}(z) J_{1-\nu}(z) + J_{-\nu}(z) J_{\nu-1}(z) = 2\sin(\nu\pi)/(z\pi),
\ee
\be
I_{\nu}(z) I_{1-\nu}(z) - I_{-\nu}(z) I_{\nu-1}(z) = -2\sin(\nu\pi)/(z\pi).
\ee

Using Eqs. (\ref{vepBes}) and (\ref{A-B-}), one can write down formula (\ref{Etvac}) for
the mean energy ratio ${\cal R}(t) = {\cal E}(t)/{\cal E}(-\tau)$ 
as follows:
\be
{\cal R}(t<0) 
= \frac{1}{8} \left[\frac{g\pi}{\sin(\nu\pi)}\right]^2\left|\frac{t}{\tau}\right|^{n+1}
 \left[K_{-}(g)K_{+}(y) + K_{+}(g)K_{-}(y) - 2 K_{0}(g)K_{0}(y)  \right],
\label{RKKK}
 \ee
\be
{\cal R}(t>0) 
= \frac{1}{8} \left[\frac{g\pi}{\sin(\nu\pi)}\right]^2\left|\frac{t}{\tau}\right|^{n+1}
 \left[K_{-}(g)\tilde{K}_{+}(y) + K_{+}(g)\tilde{K}_{-}(y) - 2 K_{0}(g)\tilde{K}_{0}(y)  \right],
\label{RKKK+}
 \ee
where
\[
K_{+}(z) =  J_{\nu-1}^2(z) + J_{\nu}^2(z) , \qquad K_{-}(z) =  J_{1-\nu}^2(z) + J_{-\nu}^2(z) , 
\]
\[
\tilde{K}_{+}(z) =   I_{\nu-1}^2(z) - I_{\nu}^2(z), \qquad 
\tilde{K}_{-}(z) =   I_{1-\nu}^2(z) - I_{-\nu}^2(z) , 
\]
\[
K_0(z) = J_{\nu-1}(z) J_{1-\nu}(z) - J_{\nu}(z) J_{-\nu}(z).
\]
\[
\tilde{K}_0(z) = I_{\nu-1}(z) I_{1-\nu}(z) - I_{\nu}(z) I_{-\nu}(z).
\]

\subsection{Adiabatic evolution of energy at $t<0$}

If $t<0$, the known leading term of the asymptotic formula for the Bessel functions of 
large arguments \cite{Grad,BE},
\be
J_{\nu}(z) \sim \sqrt{\frac{2}{\pi z}}\cos\left(z - \frac{\nu\pi}{2} - \frac{\pi}{4}\right),
\label{asJ}
\ee
results in the following simple expressions at $z \gg 1$
(with corrections of the order of $z^{-2}$):
\be
K_{\pm}(z) = \frac{2}{\pi z}, \qquad K_0(z) = \frac{2 \cos(\nu\pi)}{\pi z}.
\label{KPM)}
\ee
Hence, in the adiabatic limit ($g\gg 1$ and $y \gg 1$), we obtain 
\be
{\cal R}(t<0) 
= \frac{\omega(t) \left[1 - \cos^2(\nu\pi)\right]}{\omega_0\sin^2(\nu\pi)} = \frac{\omega(t)}{\omega_0}.
 \label{adius}
\ee
This adiabatic formula holds for any value of the power $n$ if $t<0$. 

In view of formula (\ref{Jnuy0}),
the only nonzero contribution to the right-hand side of Eq. (\ref{RKKK}) at $t \to 0$
is given by the function 
$K_{+}(y) \approx J_{\nu-1}^2(y) \approx 
\left[ (y/2)^{\nu-1}/\Gamma(\nu)\right]^2 \sim |t|^{-(n+1)}$,
whereas  the contributions of $K_{-}(y) \sim |t|^{-1}$ and $K_{0}(y) \sim |t|^0$ are eliminated
by the term $|t|^{(n+1)}$. Hence,
\be
{\cal R}(t=0) = \frac{\pi g^{2\nu-1}}{\left[ 2^{\nu} \Gamma(\nu) \sin(\pi\nu)\right]^2}, \qquad g \gg 1.
\ee
This means, in particular, that the adiabatic formula (\ref{adius}) holds under the condition
$\omega(t)/\omega_0 \gg g^{2\nu-1}$ (provided $g =2\nu \omega_0 \tau \gg 1$).
Remember that $2\nu -1 <0$.

\subsection{The harmonic oscillator revival}

The case when function $\gamma(t)$ returns to positive values $\omega^2(t)$ for $t>0$ (when it is given
by the expression in the second line of Eq. (\ref{gt}) with the positive sign) was studied in Ref.
\cite{aditrip}. In this case, the energy ratio in the adiabatic regime is given by the formula
\be
{\cal R}(t>0) 
= \frac{\omega(t) \left[1 + \cos^2(\nu\pi)\right]}{\omega_0\sin^2(\nu\pi)}.
 \label{adius+}
\ee
Constant coefficients $u_{\pm}$ depend on parameter $\nu$ as follows
\[
u_{+} = [\sin(\nu\pi)]^{-1}, \qquad u_{-} = i\cot(\nu\pi).
\]
Consequently, the squeezing coefficient satisfies the inequalities
\[
\tan^2(\nu\pi/2)  \le s_t/s_{-\tau} \le \cot^2(\nu\pi/2).
\]


\section{Evolution of the inverted oscillator energy}

\subsection{Asymptotic evolution with the power function $\gamma(t)$}

If the oscillator becomes inverted with a slowly varying function $\gamma(t)$, 
given by Eq. (\ref{gt}) under the condition $\omega_0 \tau \gg 1$,
we need asymptotic formulas 
for the modified Bessel functions of large arguments.
Remembering that function $I_{\nu}(z)$ is real for real values of its argument, we take the 
arithmetic average value (or the real part) between two equivalent expressions
in formula 8.451.5 from the reference book \cite{Grad}:  
\be
I_{\nu}(z) \sim \frac{e^z}{(2\pi z)^{1/2}} 
\sum_{k=0}^{\infty} \frac{\Gamma(\nu + k + 1/2)}{(2z)^k k! \Gamma(\nu - k + 1/2)}
\left[ (-1)^k - \sin(\nu\pi) e^{-2z }\right].
\label{asympI}
\ee
Taking into account only terms with $k=0$, one would obtain slowly varying (and incorrect) 
expressions for the functions $\tilde{K}_{\pm}(z)$ and $\tilde{K}_{0}(z)$, of the order of $z^{-1}$.
To obtain correct asymptotic expressions for these functions, one has to 
take into account the higher order terms with $k=1$ in Eq. (\ref{asympI}).
Then, the exponentially small corrections containing $e^{-2z}$ can be neglected if $z\gg1$. 
The correct expressions are as follows,
\be
\tilde{K}_{\pm}(z) \approx \tilde{K}_0(z) \approx \frac{2\nu-1}{2\pi z^2}e^{2z}.
\ee
Hence, the absolute value of the mean energy grows exponentially in the asymptotic adiabatic regime:
\be
{\cal R}(t>0) \approx \frac{(2\nu-1) \exp[2y(t)]}{8g(t/\tau)\cos^2(\nu\pi/2)},
\quad g \gg 1, \quad y(t) = g \left(\frac{t}{\tau}\right)^{1+ n/2}   \gg 1.
\ee
Since $2\nu-1 <0$, this asymptotic time dependent ratio is always negative (for any value of parameter $n$).

\subsection{If function $\gamma(t)$ becomes negative and constant} 

If $\gamma(t)= -\kappa^2$ with $\kappa =const >0$ for $t>t_* >0$, 
Eq. (\ref{eq}) has an exact solution:
\be
\vep(t) = [2\kappa]^{-1/2} \left[ v_{+} e^{\kappa t} + 
 v_{-} e^{-\kappa t} \right], 
\qquad \dot\vep(t) = [\kappa/2]^{1/2} 
\left[ v_{+} e^{\kappa t} -  v_{-} e^{-\kappa t} \right].
\label{adsolinv}
\ee
In view of Eq. (\ref{Wr}), constant complex coefficients $v_{\pm}$ must satisfy the condition  
\be
\mbox{Im}\left(v_{+}v_{-}^*\right) =1.
\label{vpmcond}
\ee
In this case, the time-independent mean energy equals
\be
\langle {\cal E}\rangle(t>t_*) = -\kappa \left[ 
\langle x^2\rangle_{-\tau}\,\omega_0 \mbox{Re}\left(v_{+}\right) \mbox{Re}\left(v_{-}\right)
+ \frac{\langle p^2\rangle_{-\tau}}{\omega_0} \mbox{Im}\left(v_{+}\right) \mbox{Im}\left(v_{-}\right)
+ \frac12 \langle px + xp\rangle_{-\tau}\, \mbox{Im}\left(v_{+} v_{-}\right)\right].
\label{E-inv}
\ee
All the second-order moments grow unlimitedly with time. 
Under the conditions (\ref{special}), we can write
\[
\langle x^2\rangle_t =  \langle x^2\rangle_{-\tau} \frac{\omega_0}{2\kappa}
\left[ |v_{+}|^2 e^{2\kappa t} + |v_{-}|^2 e^{-2\kappa t} + 2\mbox{Re} \left(v_{+}v_{-}^*  \right)\right],
 \]
\[
\langle p^2 \rangle_t  = \langle p^2\rangle_{-\tau}  \frac{\kappa}{2\omega_0} 
\left[ |v_{+}|^2 e^{2\kappa t} + |v_{-}|^2 e^{-2\kappa t} - 2\mbox{Re} \left(v_{+}v_{-}^*  \right)\right],
 \]
\[
\langle xp + px \rangle_t = \langle x^2\rangle_{-\tau}\, \omega_0 
\left[ |v_{+}|^2 e^{2\kappa t} - |v_{-}|^2 e^{-2\kappa t}\right].
\]
Nonetheless, the exponentially increasing terms
 are canceled in the formula for the mean energy, which does not depend on time if $\kappa = const$.
For the initial $N$th Fock state, we obtain
\[
\langle {\cal E}\rangle = -\kappa \mbox{Re} \left(v_{+}v_{-}^*  \right)(N+1/2).
\]

\subsection{Instantaneous jump to the inverted oscillator}

The coefficients $v_{\pm}$ can be easily calculated in the case of
a sudden jump of function $\gamma(t)$ at $t=0$ from the value $\omega_0^2$ 
to the constant value $-\kappa^2$, using the conditions of continuity 
of functions $\vep(t)$ and $\dot{\vep}(t)$:
\be
v_{\pm} = \frac{1}{\sqrt{2}}\left(\sqrt{\frac{\kappa}{\omega_0}} \pm i \sqrt{\frac{\omega_0}{\kappa}}
\right), \qquad
v_{+}v_{-}^* = i + \frac12\left(\frac{\kappa}{\omega_0} - \frac{\omega_0}{\kappa}\right).
\label{jump}
\ee
Then, formula (\ref{E-inv}) yields the following mean energy after the jump for any initial state:
\be
\langle {\cal E}\rangle_{t>0} = \frac{1}{2} \left[ 
\langle p^2\rangle_{-\tau} - \kappa^2 \langle x^2\rangle_{-\tau}\right].
\label{E-inv-jump}
\ee
This result can seem obvious, because the wave function and its mean values do not change during
the instantaneous transformation of the Hamiltonian. However, it is interesting, because
the mean energy after the jump can assume any value, depending on the ratio $\kappa/\omega_0$
and initial conditions. In particular, under the initial conditions (\ref{special}), 
the mean energy turns into zero if $\kappa = \omega_0$.
For the initial $N$-th Fock state we have
\be
\langle {\cal E}\rangle_{t>0}^{Fock} = \frac{\hbar\omega_0}{4} \left(2N+1 \right) 
\left( 1 - \frac{\kappa^2}{\omega_0^2} \right).
\label{E-inv-jump-Fock}
\ee

\section{Energy fluctuations}
\label{sec-fluct}

It can be interesting to know the strength of the {\em energy fluctuations\/} after the
frequency passes through zero.
These fluctuations can be characterized by the variance 
$\sigma_E = \langle E^2\rangle - \langle E\rangle^2$. 
Using the solutions (\ref{solxp}) of the Heisenberg equations of motion, one can write $\sigma_E$
in terms of the initial fourth- and second-order moments of the canonical variables $x$ and $p$ and 
various products of functions $\vep(t)$, $\dot\vep(t)$ and their complex conjugated partners.
The complete formula is rather cumbersome in the most general case. For this reason,
we consider here (following \cite{aditrip}) the case of the initial Fock quantum state $|N\rangle$. 
In this special case (as well as for arbitrary diagonal mixtures of the Fock states), 
the nonzero statistical moments are those containing {\em even powers\/} of each variable,
$x$ or $p$. After some algebra, one can obtain the following formula (using dimensionless
variables and assuming $\hbar=m=\omega_0 =1$, 
so that $\langle x^4 \rangle = \langle p^4\rangle$) \cite{aditrip}:
\[
16\langle E^2\rangle_t = 2\langle x^4 \rangle_{-\tau} \left(A^2 + B^2\right)
+ \langle x^2 p^2 + p^2 x^2\rangle_{-\tau} \left(A^2 - B^2\right)
+ \langle (x p + p x)^2\rangle_{-\tau} C^2,
\]
where
\[
A(t) = \gamma(t)|\vep(t)|^2 +|\dot\vep(t)|^2, \qquad
B(t) = \mbox{Re}\left[\gamma(t)\vep^2(t) + {\dot\vep}^2(t) \right], 
\qquad
C(t) = \mbox{Im}\left[\gamma(t)\vep^2(t) + {\dot\vep}^2(t) \right].
\]

\subsection{The case of harmonic oscillator at $t>0$}

In the adiabatic regime (\ref{adsol+}) we have
\be
A = 2\omega(t)\left(u_{+}|^2 +|u_{-}|^2\right), \qquad
B = 4\omega(t) \mbox{Re}\left(u_{+}u_{-}\right), \qquad
C = 4\omega(t) \mbox{Im}\left(u_{+}u_{-}\right).
\label{ABC}
\ee
For the initial Fock state $|N\rangle$ we have
\[
\langle x^4 \rangle_{-\tau} = \frac34 \left( 2N^2 + 2N +1 \right), \qquad
\langle x^2 p^2 + p^2 x^2\rangle_{-\tau} = \frac12 \left( 2N^2 + 2N -1 \right), 
\]
\[
\langle (x p + p x)^2\rangle_{-\tau} = 2 \left( N^2 + N +1 \right).
\]
Hence, 
\[
\langle E^2\rangle_t/\omega^2(t) = \left(u_{+}|^2 +|u_{-}|^2\right)^2 (N+1/2)^2
+ 2|u_{+}u_{-}|^2 \left( N^2 + N +1 \right).
\]
Remembering that the mean energy equals 
$\langle E\rangle_t = \omega(t) \left(u_{+}|^2 +|u_{-}|^2\right) (N+1/2)$,
we arrive at the following formula for the energy variance:
\be
\sigma_E(t) = 2 \omega^2(t) |u_{+}u_{-}|^2 \left( N^2 + N +1 \right), \qquad
\frac{\sigma_E(t)}{\langle E\rangle^2_t} = 2 |u_{+}u_{-}|^2 \frac{N^2 + N +1}{N^2 + N +1/4}.
\ee
In the absence of zero frequency values we have $u_{-} =0$. In this case, $\sigma_E(t) \equiv 0$,
in accordance with the Born--Fock adiabatic theorem. However, this theorem is broken when the frequency
passes through zero value. 
For  the initial vacuum state ($N=0$) and the power index $n=2$ of the single frequency
transition through zero value, we obtain 
${\sigma_E(t)}/{\langle E\rangle^2_t} =16$. This ratio can be four times smaller if $N\gg 1$.

\subsection{The case of inverted oscillator}

In the case of inverted oscillator with a constant parameter $\kappa$, 
one should replace $\omega(t)$ with $i\kappa$ and use
the solution (\ref{adsolinv})  instead of (\ref{adsol+}).
Then, we have the following expressions instead of (\ref{ABC}):
\be
A= -2\kappa\mbox{Re}\left(v_{+}v_{-}^*\right), \qquad
B= -2\kappa\mbox{Re}\left(v_{+}v_{-}\right), \qquad
C= -2\kappa\mbox{Im}\left(v_{+}v_{-}\right).
\label{ABC-inv}
\ee
In the case of a sudden jump (\ref{jump}), we obtain
\[
A = \omega_0\left(1 - \frac{\kappa^2}{\omega_0^2} \right), \qquad
B = -\omega_0\left(1 + \frac{\kappa^2}{\omega_0^2} \right), \qquad C = 0.
\]
Hence, 
\be
\langle E^2\rangle_{t>0}^{Fock} = \frac{1}{16} (\hbar\omega_0)^2 \left[
 3\left( 2N^2 + 2N +1 \right) \left(1 + \frac{\kappa^4}{\omega_0^4} \right)
- 2 \frac{\kappa^2}{\omega_0^2} \left( 2N^2 + 2N -1 \right) \right],
\label{E2inv-jump}
\ee
\be
\sigma_{E}^{Fock}(t>0) = \frac{1}{8} (\hbar\omega_0)^2  \left( N^2 + N +1 \right)
\left(1 + \frac{\kappa^2}{\omega_0^2} \right)^2 .
\label{sigE-jump}
\ee

\section
{Probability density distribution over   energy eigenstates
of the inverted oscillator}
\label{sec-probdistr}

It is interesting to calculate the probability density
of measuring the energy eigenvalues of the inverted oscillator with a fixed parameter $\kappa$.
The evolution of the initial Fock state $|n\rangle$ is determined completely by the function $\vep(t)$
introduced in Sec. \ref{sec-quantgen}.
The time dependent wave function has the following form \cite{Husimi,PP,MMT70}
(we use dimensionless units with $\hbar=m =1$ in this section):
\be
\psi_n(x,t) = \left( n!\,\vep \sqrt{\pi}\right)^{-1/2}
\left(\frac{\vep^*}{2\vep}\right)^{n/2}
\exp\left( \frac{i\dot\vep}{2\vep} x^2\right) H_n\left(\frac{x}{|\vep|}\right),
\label{Nxt}
\ee
where $H_n(z)$ is the Hermite polynomial.
The wave functions $\psi(x;E)$ of energy eigenstates are solutions of 
the stationary Schr\"odinger equation
\be
d^2 \psi/dx^2 + \left(\kappa^2 x^2 + 2E\right)\psi = 0.
\label{SEE}
\ee
It is known that the energy spectrum of the inverted oscillator is continuous and doubly degenerate, with
$-\infty < E < \infty$. Two independent normalized solutions to Eq. (\ref{SEE}) can be found in papers 
\cite{Chrus04,Ullinger22} (we use a slightly different notation):
\be
\psi_{\pm}(x;E) = {\cal N} D_{\mu}\left(\mp x \sqrt{-2i\kappa}\right),
\label{solD}
\ee
where $D_{\mu}(z)$ is the parabolic cylinder function defined as in books \cite{Grad,BE}.
The index $\mu$ and the normalization constant ${\cal N}$ are related to the normalized energy 
$\tilde{E} = E/\kappa$ as follows (we assume $\kappa >0$):
\be
\mu = -\,\frac12 + i\tilde{E}, \qquad 
{\cal N} = \frac{\Gamma(-\mu) \exp(\pi \tilde{E}/4)}{\pi(8\kappa)^{1/4}},
\label{muN}
\ee
where $\Gamma(z)$ is the Euler gamma function. With this normalization, functions (\ref{solD}) obey the relations
\be
\int_{-\infty}^{\infty} \psi^*_{\pm}(x;E) \psi_{\pm}(x;E^{\prime}) dx = \delta(E-E^{\prime}).
\label{delta-E}
\ee

In view of the double degeneracy of energy levels in the inverted regime,
the instantaneous energy probability distribution 
with respect to the normalized energy
can be written as follows,
\be
P_n(\tilde{E}) = \kappa \left|\int_{-\infty}^{\infty} \psi_n^*(x,t)|\psi_{+}(x;E) dx \right|^2
+ \kappa \left|\int_{-\infty}^{\infty} \psi_n^*(x,t)|\psi_{-}(x;E) dx \right|^2.
\label{defPn}
\ee
Hence, one has to calculate the integrals of the following structure:
\be
I_{\mu,n}(a,b,c) = \int_{-\infty}^{\infty} e^{-a x^2} D_{\mu}(cx)H_n(bx) dx .
\label{DHint}
\ee
Although explicit expressions for the integral (\ref{DHint}) are absent 
in the main reference books, such as \cite{Grad,BE,Br2}, it can be calculated analytically
with the aid of the integral representation of the parabolic cylinder function
(see formula 9.241.2 from \cite{Grad}),
\be
D_{\mu}(z) = \frac{e^{-z^2/4}}{\Gamma(-\mu)} \int_0^{\infty} \exp\left(-zy - y^2/2\right)
y^{-\mu -1} dy.
\label{Dmuint}
\ee
Putting the integral (\ref{Dmuint}) in Eq. (\ref{DHint}) and changing the order of integrations,
we arrive at the integral
\[
\int_{-\infty}^{\infty} \exp\left[-(a+c^2/4)x^2 -ycx\right] H_n(bx) dx,
\]
which can be calculated with the aid of formula 
7.374.8 from \cite{Grad},
\be
\int_{-\infty}^{\infty} e^{-(\xi -\eta)^2} H_n(\alpha\xi) d\xi =
\sqrt{\pi} \left(1 - \alpha^2\right)^{n/2} H_n\left(\frac{\alpha\eta}{\sqrt{1 - \alpha^2}}\right),
\label{intexpHn}
\ee
with
\[
\alpha = \frac{b}{\sqrt{a+c^2/4}}, \qquad \eta = -\,\frac{yc}{2\sqrt{a+c^2/4}}.
\]
Hence, we can write
\be
I_{\mu,n} = \frac{(-1)^n \sqrt{\pi} \left(a+c^2/4 -b^2\right)^{n/2}}
{\Gamma(-\mu) \left(a+c^2/4 \right)^{(n+1)/2}}
\int_0^{\infty} \exp\left(- B y^2\right) H_n(Ay)
y^{-\mu -1} dy,
\label{intAB}
\ee
where
\[
A = \frac{bc/2}{\sqrt{(a+c^2/4)(a+c^2/4-b^2)}}, \qquad B = \frac{a-c^2/4}{2(a+c^2/4)}.
\]
The value of the integral in (\ref{intAB}) depends on the parity of the Hermite polynomials,
according to formulas 7.376.2 and 7.376.3 from \cite{Grad}:
\[
\int_0^{\infty} e^{-2\gamma x^2} x^{\nu} H_{2k}(x) dx = (-1)^k 2^{2k -\frac32 -\frac{\nu}{2}}
\frac{\Gamma\left(\frac{\nu+1}{2}\right) \Gamma\left(k+\frac{1}{2}\right)}
{\sqrt{\pi} \gamma^{\frac{\nu+1}{2}}} F\left(-k,\frac{\nu+1}{2}; \frac12; \frac{1}{2\gamma}\right),
\]
\[
\int_0^{\infty} e^{-2\gamma x^2} x^{\nu} H_{2k+1}(x) dx = (-1)^k 2^{2k  -\frac{\nu}{2}}
\frac{\Gamma\left(\frac{\nu}{2} +1\right) \Gamma\left(k+\frac{3}{2}\right)}
{\sqrt{\pi} \gamma^{\frac{\nu}{2}+1}} F\left(-k,\frac{\nu}{2} +1; \frac32; \frac{1}{2\gamma}\right),
\]
where $F(a,b;c;z)$ is the Gauss hypergeometric function.
In our case, we have
\[
\nu= -\mu -1 = -\frac12 -i\tilde{E}, \qquad 
2\gamma = {B}/{A^2} = 2\left(a-c^2/4 \right)\left(a+c^2/4-b^2\right)/(bc)^2.
\]
If $\vep(t)$ is given by Eq. (\ref{adsolinv}), we have
\[
a= \frac{i \dot{\vep}^*}{2\vep^*} = \frac{i\kappa}{2}
 \frac{v_{+}^* e^{\kappa t} - v_{-}^* e^{-\kappa t} }{v_{+}^* e^{\kappa t} + v_{-}^* e^{-\kappa t}},
\qquad
b= 1/|\vep| = \frac{\sqrt{2\kappa}}{\left|v_+ e^{\kappa t} + v_- e^{-\kappa t}\right|}, \qquad
c^2/4 = -i\kappa/2,
\]
\[
a-c^2/4 = \frac{i\kappa v_{+}^* e^{\kappa t}}{v_{+}^* e^{\kappa t} + v_{-}^* e^{-\kappa t}}, \qquad
a+c^2/4 = -\,\frac{i\kappa v_{-}^* e^{-\kappa t}}{v_{+}^* e^{\kappa t} + v_{-}^* e^{-\kappa t}}.
\]

Hereafter, we consider the special case of a sudden jump, with coefficients (\ref{jump}).
 Since the energy distribution cannot depend on time for 
the stationary Hamiltonian at $t>0$,
the coefficients $a$ and $b$ can be taken at the initial instant $t=0+$.
Hence, we can use more simple expressions
\[
a = \frac{i\kappa}{2}
 \frac{v_{+}^*  - v_{-}^*  }{v_{+}^*  + v_{-}^* } = \omega_0/2, \qquad 
b = \frac{\sqrt{2\kappa}}{\left|v_+  + v_- \right|} =\sqrt{\omega_0},
\]
\[
a \pm c^2/4 = \frac12\left(\omega_0 \mp i\kappa\right), \qquad
a + c^2/4 - b^2 = -\,\frac12\left(\omega_0 + i\kappa\right),
\]
so that
\[
2\gamma = \frac{\left(\omega_0 + i\kappa\right)^2}{4i\kappa\omega_0} =
\frac{\left(1 + i\rho\right)^2}{4i\rho}, \qquad \rho \equiv \kappa/\omega_0.
\]
Final formulas for the energy probability densities after the jump are as follows,
\be
P_{2k}(\tilde{E};\rho) = \frac{\Gamma(k+\frac12) \rho^{1/2}}{2 \pi^2 k! (1 +\rho^2)^{1/2}}
e^{ 2\tilde{E}\Phi(\rho)} 
\left|\Gamma\left(\frac14 - i\frac{\tilde{E}}{2}\right)
F\left(-k, \frac14 - i\frac{\tilde{E}}{2}; \frac12; z(\rho) \right)\right|^2 ,
\label{P2k-jump}
\ee
\be
P_{2k+1}(\tilde{E};\rho) = \frac{ 8\Gamma(k+\frac32) \rho^{3/2}}{ \pi^2 k! (1 +\rho^2)^{3/2}}
e^{ 2\tilde{E}\Phi(\rho)} 
\left|\Gamma\left(\frac34 - i\frac{\tilde{E}}{2}\right)
F\left(-k, \frac34 - i\frac{\tilde{E}}{2}; \frac32; z(\rho) \right)\right|^2 ,
\label{P2k+1-jump}
\ee
\[
\Phi(\rho) = \frac{\pi}{4} -\tan^{-1}(\rho) = \tan^{-1}\left(\frac{1-\rho}{1+\rho}\right), \qquad
z(\rho) = \frac{4i\rho}{(1 +i\rho)^2}.
\]

If $\rho=1$, the formulas can be simplified as follows,
\be
P_{2k}(\tilde{E}) = \frac{\Gamma(k+1/2)}{2^{3/2} \pi^2 k!}
\left|\Gamma\left(\frac14 - i\frac{\tilde{E}}{2}\right)
F\left(-k, \frac14 - i\frac{\tilde{E}}{2}; \frac12; 2\right)\right|^2,
\label{P2k}
\ee
\be
P_{2k+1}(\tilde{E}) = \frac{2^{3/2}\Gamma(k+3/2)}{ \pi^2 k!}
\left|\Gamma\left(\frac34 - i\frac{\tilde{E}}{2}\right)
F\left(-k, \frac34 - i\frac{\tilde{E}}{2}; \frac32; 2\right)\right|^2.
\label{P2k+1}
\ee
We used the known formula $\Gamma(2z) = 2^{2z-1} \pi^{-1/2} \Gamma(z) \Gamma(z+1/2)$.
All probability distributions (\ref{P2k}) and (\ref{P2k+1}) are even functions of energy, 
because the mean energy after the jump with $\rho=1$ equals zero.
In particular, for $k=0$ we have
\be
P_0(\tilde{E})  = (2\pi )^{-3/2} 
\left|\Gamma\left(\frac14 - i\frac{\tilde{E}}{2}\right)\right|^2, \qquad
P_1(\tilde{E}) 
= \left(\frac{2}{\pi^{3}}\right)^{1/2} \left|\Gamma\left(\frac34 - i\frac{\tilde{E}}{2}\right)\right|^2 .
\label{PE01}
\ee
In these special cases, the correct normalization
$
\int_{-\infty}^{\infty} P(\tilde{E})d\tilde{E} =1
$
can be checked with the aid of formula 2.2.4.1 from \cite{Br2},
\be
\int_0^{\infty} |\Gamma(a +ix)|^2 dx = 2^{-2a} \pi \Gamma(2a).
\label{intmodGam}
\ee
Using the relations
\[
(a + ix) \Gamma(a + ix) = \Gamma(a +1 + ix) , \qquad
\left(a^2 + x^2\right)|\Gamma(a + ix)|^2 = |\Gamma(a +1 + ix)|^2,
\]
we obtain the integral
\be
\int_0^{\infty} x^2|\Gamma(a + ix)|^2 dx = 2^{-2a-1} \pi a \Gamma(2a).
\label{intx2Gam}
\ee
Hence, writing 
$P(\tilde{E}) = N_a |\Gamma(a + i\tilde{E}/2)|^2$,
we obtain the following formula for the mean value $\langle \tilde{E}^2 \rangle$ with respect to the 
probability densities $P_0(\tilde{E})$ and $P_1(\tilde{E})$:
$
\langle \tilde{E}^2 \rangle = 16\pi a N_a 2^{-2a -1} \Gamma(2a)
$.
For $a=1/4$ we have $\langle \tilde{E}^2 \rangle= 1/2$, while for $a=3/4$ we have $\langle \tilde{E}^2 \rangle= 3/2$.
These values coincide exactly with those given by Eq. (\ref{E2inv-jump}) for $N=0$ and $N=1$
(provided $\kappa=\omega_0$).

For $k=1$ and $k=2$, we obtain the following expressions:
\[
P_2(\tilde{E}) 
= \left(2\pi^3\right)^{-1/2}
\tilde{E}^2\left|\Gamma\left(\frac14 - i\frac{\tilde{E}}{2}\right)\right|^2 , 
\qquad
P_3(\tilde{E}) 
= \left(\frac{2}{\pi}\right)^{3/2}
\frac{\tilde{E}^2}{3}\left|\Gamma\left(\frac34 - i\frac{\tilde{E}}{2}\right)\right|^2 .
\]
\[
P_4(\tilde{E}) = \frac{\left(1 - 2\tilde{E}^2\right)^2}{6(2\pi)^{3/2}}
\left|\Gamma\left(\frac14 - i\frac{\tilde{E}}{2}\right)\right|^2 , 
\qquad
P_5(\tilde{E}) 
= \frac{6}{5(2\pi)^{3/2}}
\left(1 - \frac23 \tilde{E}^2\right)^2\left|\Gamma\left(\frac34 - i\frac{\tilde{E}}{2}\right)\right|^2 .
\]

\begin{figure}[htb]
\vspace{-0.5 cm}
\begin{center}
\includegraphics[width=0.99\textwidth]{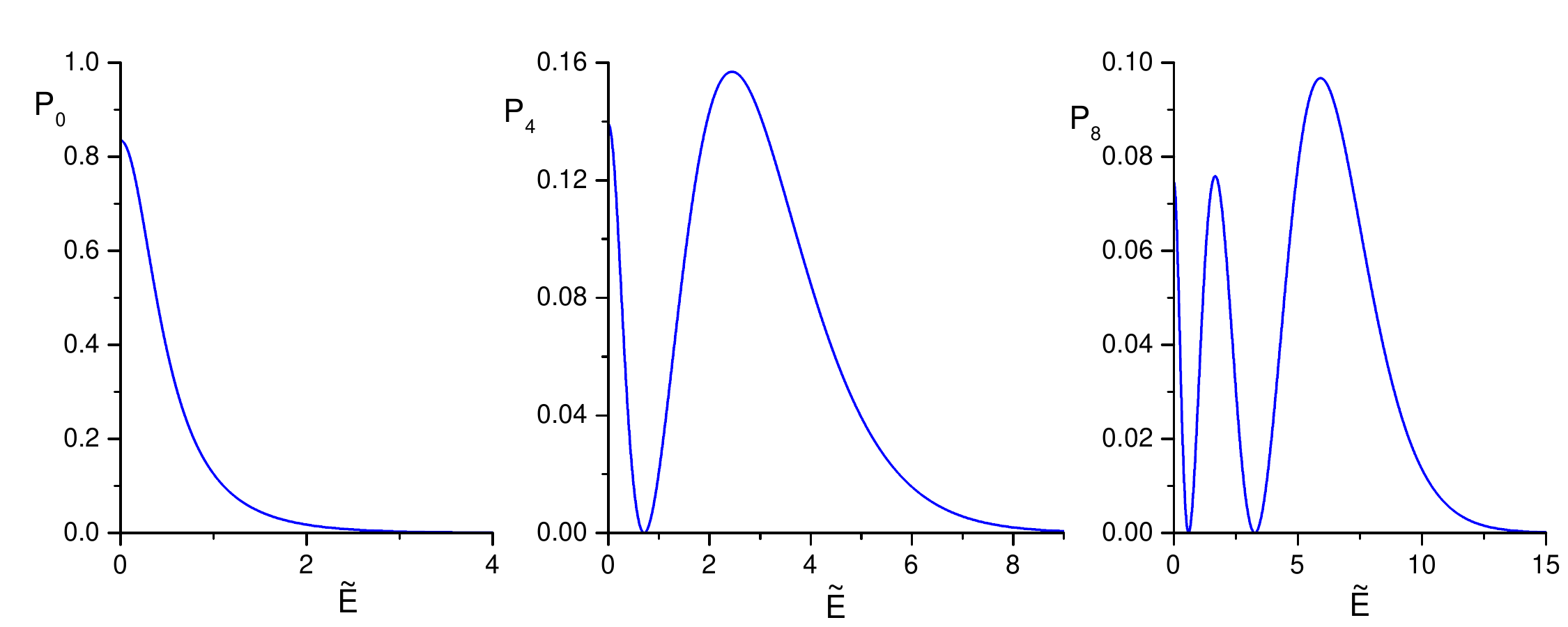} {}
\end{center}
\vspace{-0.5 cm}
\caption{Functions $P_0(\tilde{E})$, $P_4(\tilde{E})$ and $P_8(\tilde{E})$
after the jump with $\rho=1$. }
\label{fig-P048}
\end{figure}
\begin{figure}[htb]
\vspace{-0.5 cm}
\begin{center}
\includegraphics[width=0.99\textwidth]{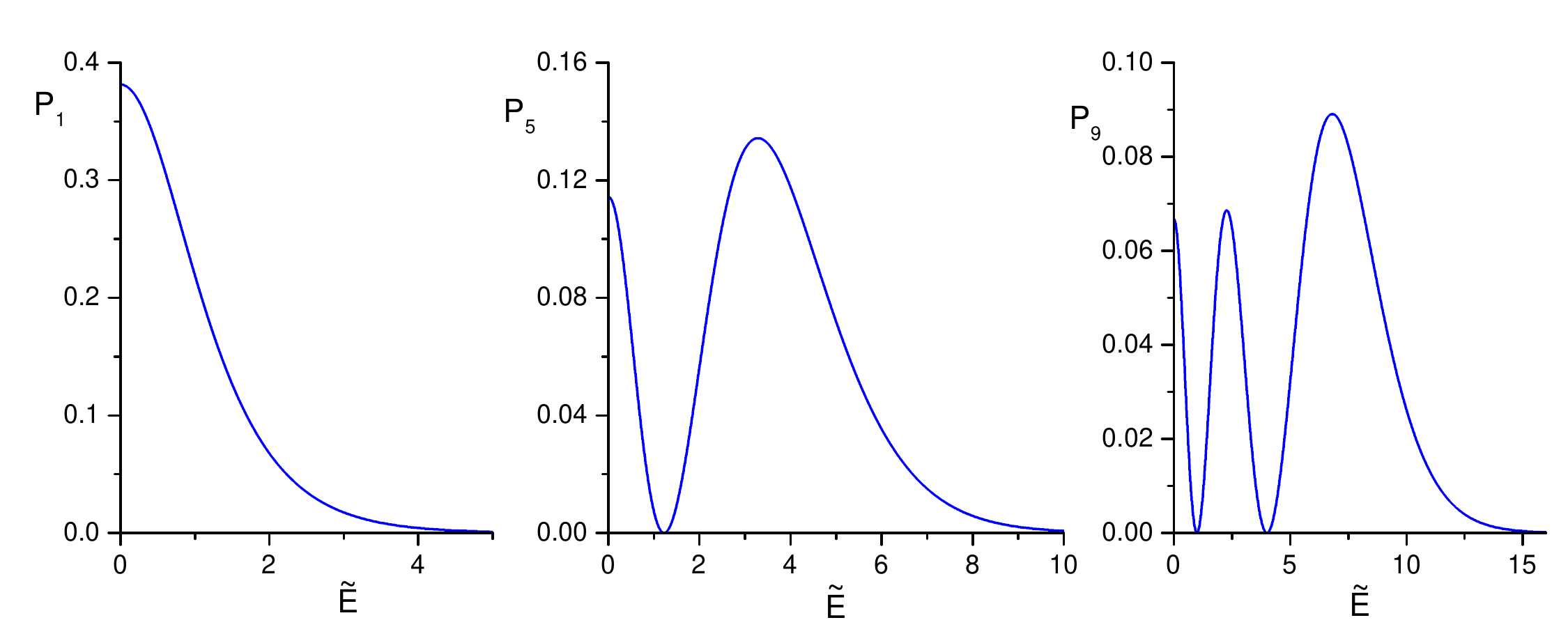} {}
\end{center}
\vspace{-0.5 cm}
\caption{Functions $P_1(\tilde{E})$, $P_5(\tilde{E})$ and $P_9(\tilde{E})$
after the jump with $\rho=1$. }
\label{fig-P159}
\end{figure}
\begin{figure}[htb]
\vspace{-0.5 cm}
\begin{center}
\includegraphics[width=0.99\textwidth]{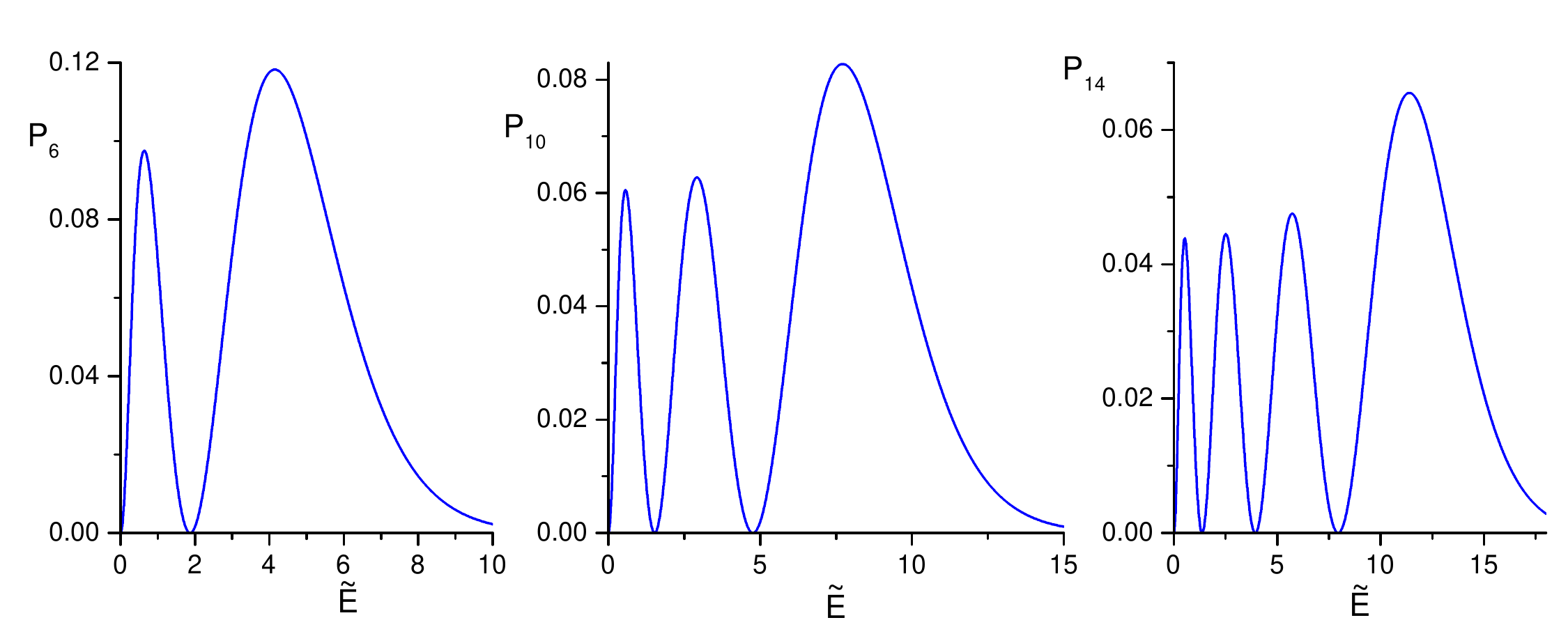} {}
\end{center}
\vspace{-0.5 cm}
\caption{Functions $P_{6}(\tilde{E})$, $P_{10}(\tilde{E})$ and $P_{14}(\tilde{E})$
after the jump with $\rho=1$. }
\label{fig-P61014}
\end{figure}
\begin{figure}[htb]
\vspace{-0.5 cm}
\begin{center}
\includegraphics[width=0.99\textwidth]{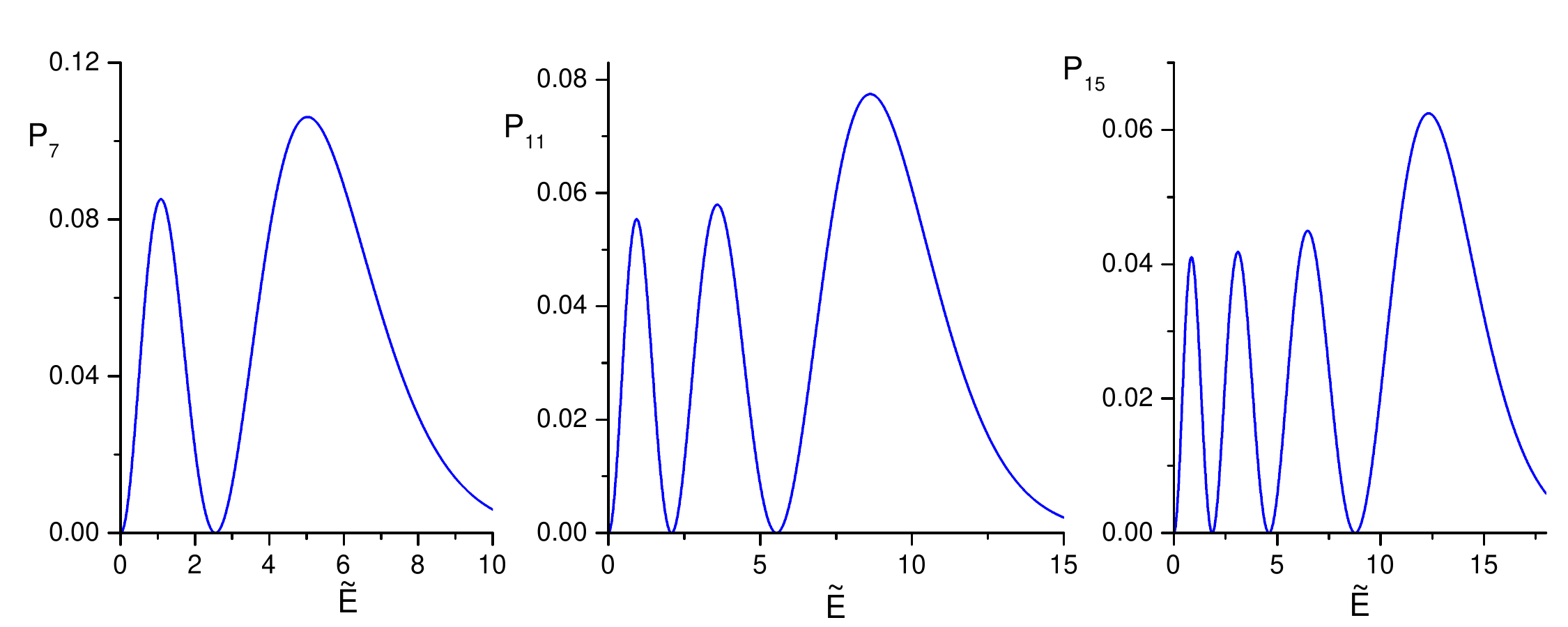} {}
\end{center}
\vspace{-0.5 cm}
\caption{Functions $P_{7}(\tilde{E})$, $P_{11}(\tilde{E})$ and $P_{15}(\tilde{E})$
after the jump with $\rho=1$. }
\label{fig-P71115}
\end{figure}

Looking at plots of functions $P_{n}(\tilde{E})$ in Figures \ref{fig-P048}--\ref{fig-P71115}
(where we consider 
the arguments $\tilde{E} \ge 0$ only, due to the symmetry of distributions), one can notice 
 that distributions $P_{2k}(\tilde{E})$ and $P_{2k+1}(\tilde{E})$
behave in a similar way, in spite of slightly different analytical expressions (\ref{P2k})
and (\ref{P2k+1}) for the initial even and odd Fock states. 
On the other hand, the parity of number $k$ turns out to be important. If this number is even
($k=0,2,4,\ldots$), the probability densities at $\tilde{E} = 0$ are different from zero.
In contrast, $P_{2k}(0) = P_{2k+1}(0)$ for odd values $k=1,3,5,\ldots$: see Figures
\ref{fig-P61014} and \ref{fig-P71115}.
All probability distributions have $[k/2]$  zeros at positive values of argument $\tilde{E}$
(and symmetric zeros at negative values). The heights of the first $[k/2]$ maximas slowly increase
with $\tilde{E}$, whereas the last maximum is significantly higher (by approximately $50\%$) and
wider. This last maximum of $P_{n}(\tilde{E})$ is attained when $\tilde{E}$ is a little smaller than $n$.
The function $P_{n}(\tilde{E})$ rapidly goes to zero at $\tilde{E} >n$.
Its asymptotic behavior  for $|\tilde{E}| > n$ can be described by the formula
$
P_n(\tilde{E}) \sim g_n(\tilde{E}) \exp(-\pi |\tilde{E}|/2)$, 
where the pre-factor $g_n(\tilde{E})$ contains some powers of $|\tilde{E}|$.
This is the consequence of formula 1.18(6) from \cite{BE}:
$
\lim_{|y|\to\infty} |\Gamma(x+iy)|^2 e^{\pi|y|} |y|^{1-2x} = 2\pi $. Numerical calculations
confirm the asymptotic formula with high accuracy.

\begin{figure}[htb]
\vspace{-0.5 cm}
\begin{center}
\includegraphics[width=0.99\textwidth]{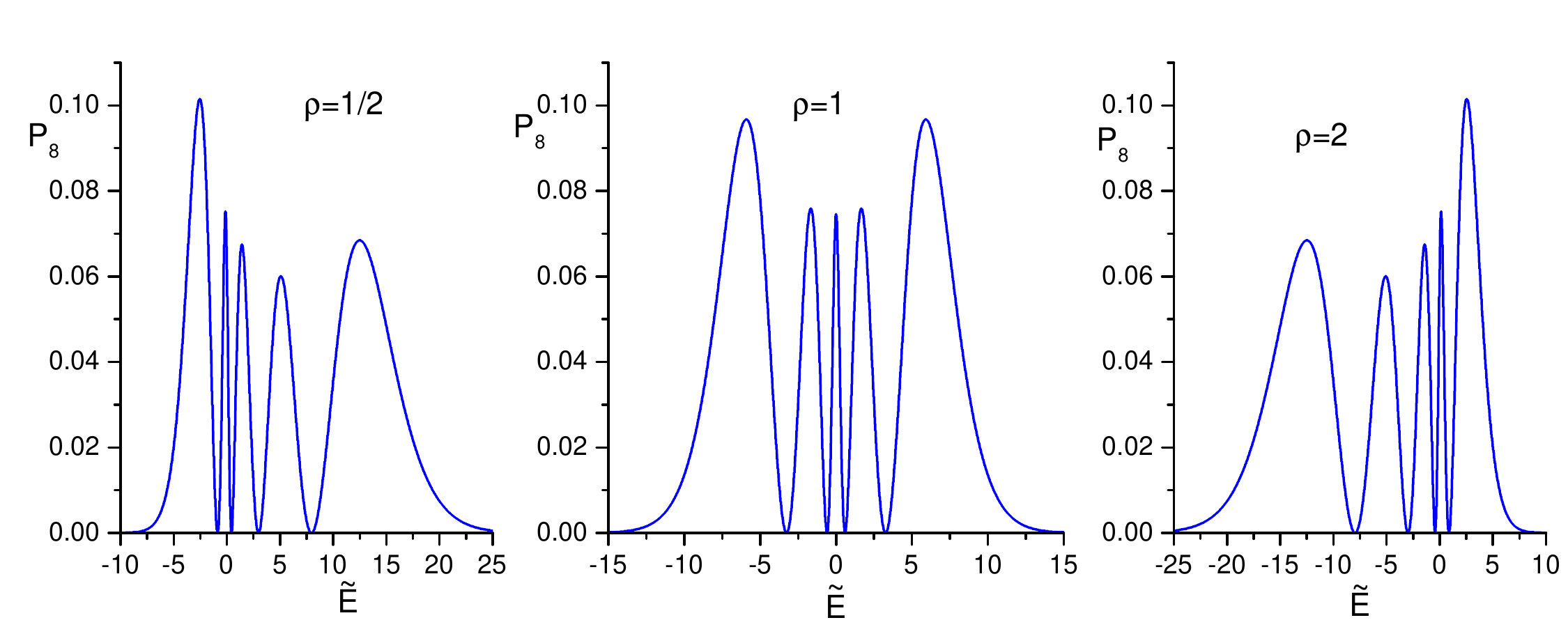} {}
\end{center}
\vspace{-0.5 cm}
\caption{Functions $P_{8}(\tilde{E})$ after the jumps with $\rho = 1/2, 1, 2$. }
\label{fig-P8jump}
\end{figure}

\begin{figure}[htb]
\vspace{-0.5 cm}
\begin{center}
\includegraphics[width=0.99\textwidth]{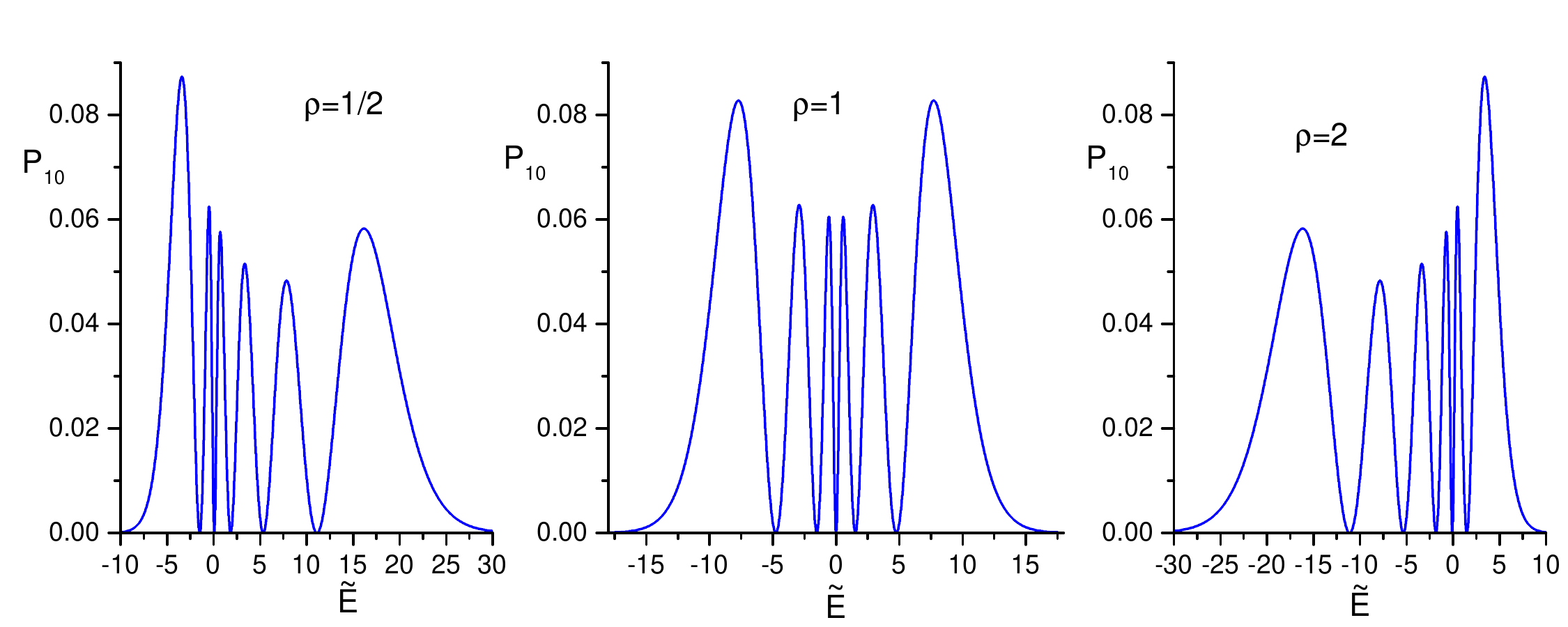} {}
\end{center}
\vspace{-0.5 cm}
\caption{Functions $P_{10}(\tilde{E})$ after the jumps with $\rho = 1/2, 1, 2$. }
\label{fig-P10jump}
\end{figure}

If $\rho \neq 1$, the 
distributions (\ref{P2k-jump}) and (\ref{P2k+1-jump}) lose the symmetry with respect
to the point $\tilde{E}=0$, in accordance with formula (\ref{E-inv-jump-Fock}). 
However, due to the relations
\[
\Phi(1/\rho) = - \Phi(\rho), \qquad
z(1/\rho) = z^*(\rho)
\]
we have the following reciprocity formula:
\be
P_n(\tilde{E};\rho) = P_n(-\tilde{E};1/\rho).
\ee
Some examples of distributions with different values of $\rho$
are shown in Figures \ref{fig-P8jump} and \ref{fig-P10jump}.

\section{Discussion}

Let us discuss briefly the main results of this work. We have obtained explicit analytical formulas
for the mean energy and its variance (characterizing the energy fluctuations) in the adiabatic
regimes after the frequency passes through zero. The behavior of energy turns out to be quite different
in two cases: when the frequency remains real and when it becomes imaginary. In the first case, the mean
energy always increases when the frequency returns to its initial value, and the increment coefficient
is determined by the exponent in the power law of the frequency crossing zero. On the other hand, if
the frequency becomes imaginary, the absolute value of mean energy increases exponentially, even in
the adiabatic regime, unless the Hamiltonian becomes time independent. It is worth emphasizing that
even small corrections to the leading terms of simple adiabatic approximate formulas are crucial
in this case, due to the unstable nature of the motion.

The mean energy does not depend on time after a sudden transformation of a harmonic oscillator
to an inverted oscillator with constant parameters. In this case, the mean energy can be both
positive and negative (or zero), depending on the ratio of the initial and final frequencies.

Comparing the results of this study with those obtained in paper \cite{aditrip}, we can expect
that the adiabatic behavior of energy and other observables can be even more intricate when
the function $\gamma(t)$ in the Hamiltonian (\ref{Ham}) passes through zero many times, changing
its signs. Probably, some kinds of quasi-chaotic behaviors could be observed. However, this
complicated problem deserves a separate study.

\section*{Acknowledgments}
This work received partial financial support from 
Brazilian funding agencies 
 National Council for Scientific and Technological Development (CNPq)
and
Fundação de Apoio à Pesquisa do Distrito Federal (FAPDF), grant number 00193-00001817/2023-43.

\end{document}